\documentclass{PoS}
\usepackage{graphicx}
\usepackage{subfigure}

\title{Charged jet spectra in proton-proton collisions with the ALICE experiment at the LHC}

\ShortTitle{Charged jet spectra in proton-proton collisions with the ALICE experiment at the LHC}

\author{\speaker{Michal VAJZER}%
        \\for the ALICE Collaboration.\\

        Nuclear Physics Institute, The Academy of Sciences of the Czech Republic\\
        \v{R}e\v{z}, Czech Republic\\
        Faculty of Nuclear Sciences and Physical Engineering, Czech Technical University in Prague\\
        Prague, Czech Republic\\
        
        E-mail: \email{michal.vajzer@cern.ch}}

\abstract{
    Jets are collimated sprays of particles resulting from fragmentation of hard scattered partons.  They are measured in different types of collisions at different energies to test perturbative Quantum Chromodynamic calculations and are used to study the hard scattering, fragmentation and hadronisation of partons.  These phenomena, measured in simple systems such as proton--proton collisions, serve as a baseline to investigate their modifications by hot and dense nuclear matter created in high energy heavy-ion collisions. 
  
    We have analysed data from minimum bias proton--proton collisions at centre of mass energy of 2.76 and 7 TeV collected using the ALICE detector system at the LHC and reconstructed the inclusive jet cross section from charged tracks at midrapidity.  We present jet spectra reconstructed using the infrared and colinear safe anti-k$_{\rm T}$~algorithm with underlying event subtraction, corrected for detector effects via unfolding for both collision energies. Furthermore, results from analyses of fragmentation distributions and jet shape observables are shown. All results are compared with measurements of other LHC experiments and with Monte Carlo generators.
}

\FullConference{The European Physical Society Conference on High Energy Physics \\ 
                18-24 July, 2013\\ 
                Stockholm, Sweden}

\begin{document}

  \section{Introduction}

    Jets are observable final states of hard scattered partons. They serve as a tool for testing of both perturbative and non-perturbative aspects of Quantum Chromodynamics (QCD), namely cross sections of scattering production and parton shower evolution. In collisions of heavy ions, due to the short formation time of hard partons, they are used as an important probe in the study of produced nuclear medium. For these studies, proton--proton collisions are an important baseline.  
    
    We present production cross sections for the charged part of jets using reconstructed charged tracks, further denoted as ``charged jets'', in proton--proton collisions with centre of mass energies of 2.76~TeV and 7~TeV, as reconstructed with the ALICE detector system. Furthermore, we show properties of these jets derived from analysis of individual jet constituents, namely average charged particle multiplicity, summed track transverse momentum in radial slices from jet axis and the momentum distribution of individual jet constituents.

  \section{Data analysis}

    \subsection{Event selection}
      We have analysed 7~TeV proton--proton collisions from the year 2010 and 2.76~TeV proton--proton collisions from 2011, recorded using A Large Ion Collider Experiment\cite{ALICE_JINST} (ALICE) at the Large Hadron Collider~(LHC). We select minimally biased events triggered with the VZERO detector which consist of two forward triggering scintillator counters. In offline analysis, the reconstructed primary vertex was required to be within 10~cm along the beam axis from the nominal interaction point. At least 2 tracks had to be associated with this primary vertex.

    \subsection{Jet reconstruction}
      Jets are the final state of hard partons, but experimentally observed only as collimated sprays of particles. In our analysis, we focus on charged particles reconstructed as tracks in the central tracking detectors, namely the Time Projection Chamber (TPC) and silicon Inner Tracking System (ITS), consisting of 2 layers of pixel, drift and strip detectors each. We obtain track distributions in full azimuth and pseudo-rapidity interval $\left| \eta\right|  <0.9$. Tracks are selected with reconstructed transverse momenta of at least 150~MeV/$c$. 
    
    \begin{figure}[t]
    \begin{center}
      \subfigure[$\sqrt{s} = 7$~TeV]{ \label{spectra7TeV} \includegraphics[width=0.47\textwidth]{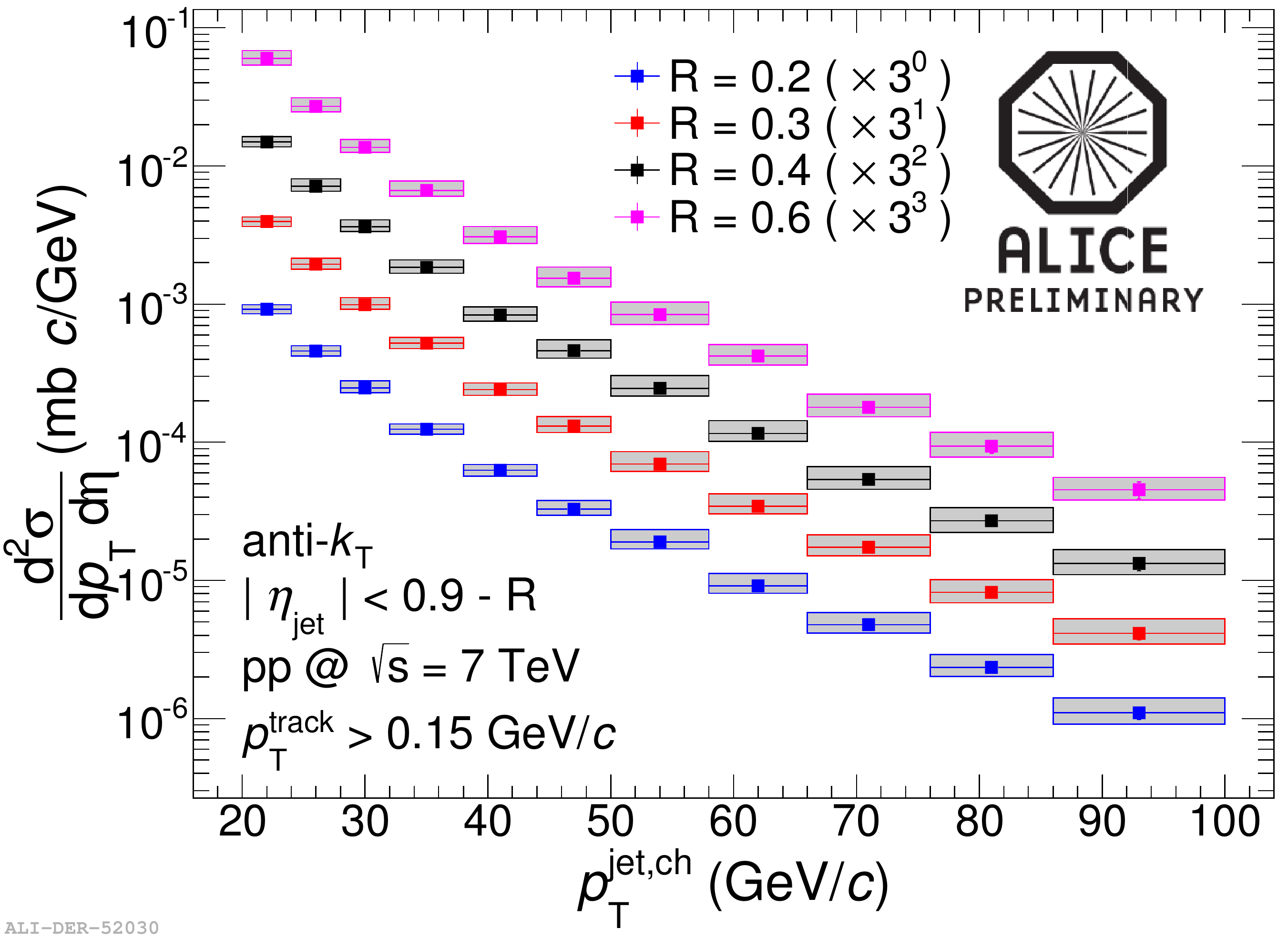} }
      \subfigure[$\sqrt{s} = 2.76$~TeV]{ \label{spectra2760GeV} \includegraphics[width=0.47\textwidth]{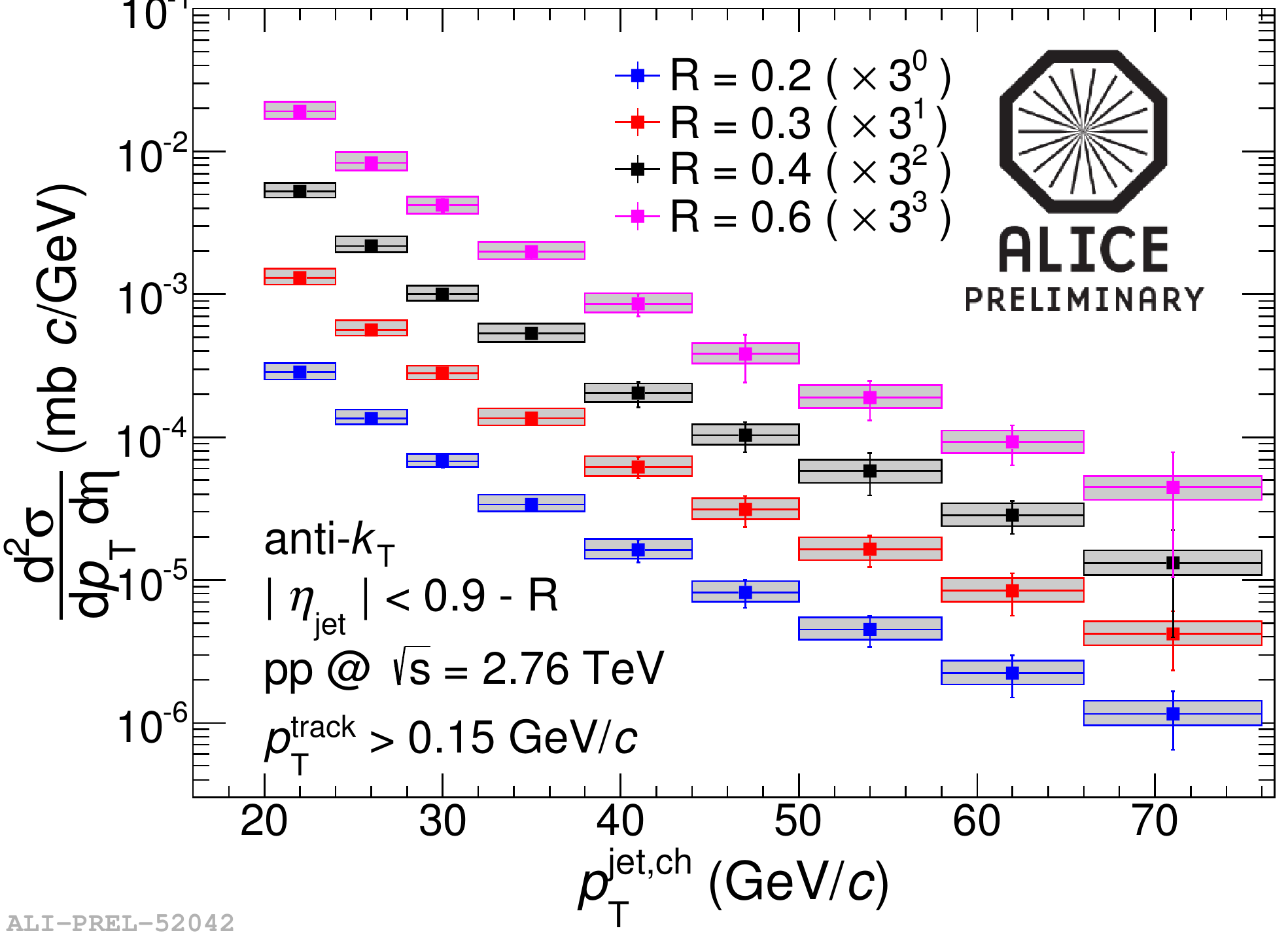} }
      \caption[Figure with jet cross sections.]{Charged anti-k$_{\rm T}$~jet spectra measured in 7~TeV and 2.76 TeV collisions, Fig.~\ref{spectra7TeV} and Fig.~\ref{spectra2760GeV} respectively. The error bars represent statistical uncertainties and boxes represent systematic uncertainties coming mainly from track reconstruction efficiency. Spectra are scaled by factors of 1, 3, 9  and 27 for resolution parameter 0.2, 0.3, 0.4 and 0.6 respectively.}
      \label{jet-spectra}
      \end{center}
    \end{figure}

    \begin{figure}[b]
      \centering
      \includegraphics[width = 0.86\textwidth]{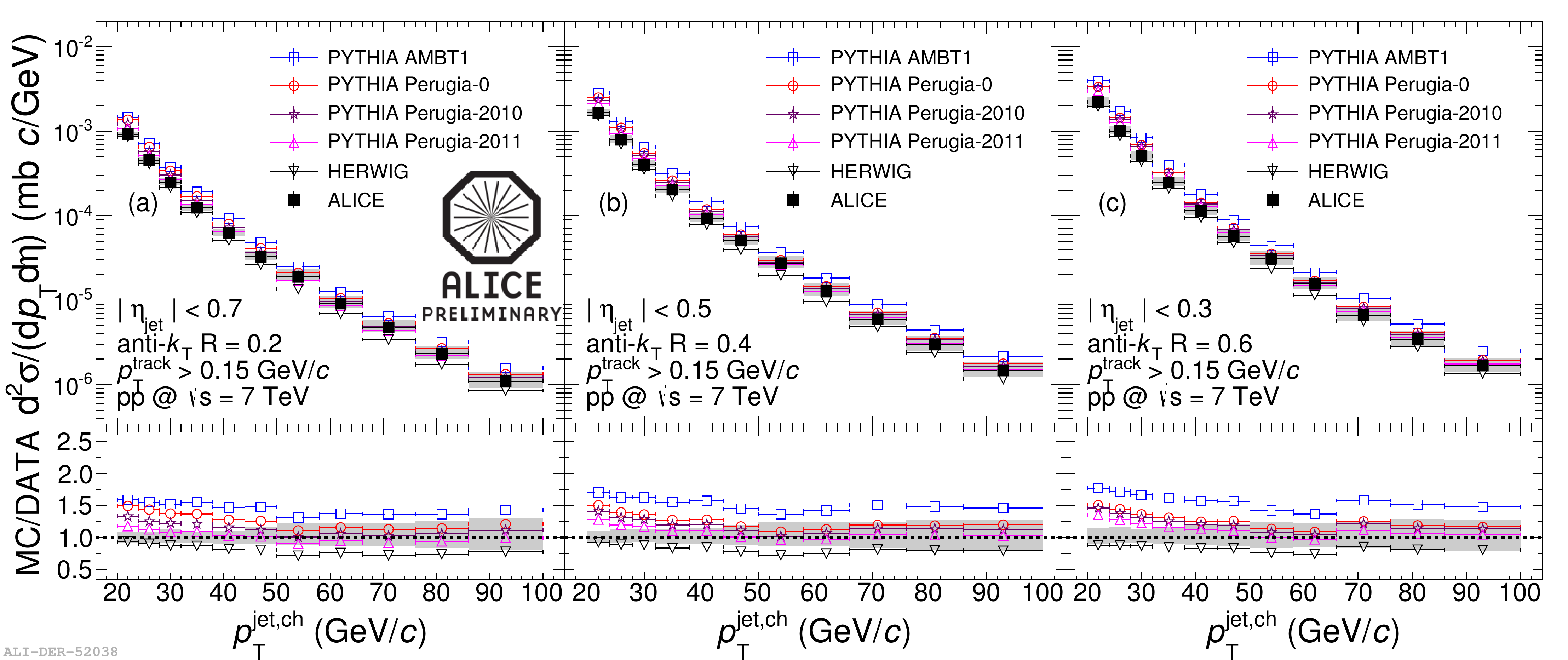}
      \caption[Figure with MC comparison of cross sections.]{Jet cross section (full black squares) measurement in proton--proton collisions at 7~TeV compared to Monte Carlo predictions obtained from PYTHIA Perugia-0 (open red circles), PYTHIA Perugia-2010 (open purple stars), PYTHIA Perugia-2011 (open magenta triangles), PYTHIA AMBT1 (open blue squares) and HERWIG (open black triangles). Lower panel of each plot shows ratios of Monte Carlo predictions to data.}
      \label{mc-comp}
    \end{figure}

    \begin{figure}[t]
      \begin{center}
        \subfigure[anti-k$_{\rm T}$,~R = 0.4 ] { \label{ATLAS-R04} \includegraphics[width=0.46\textwidth]{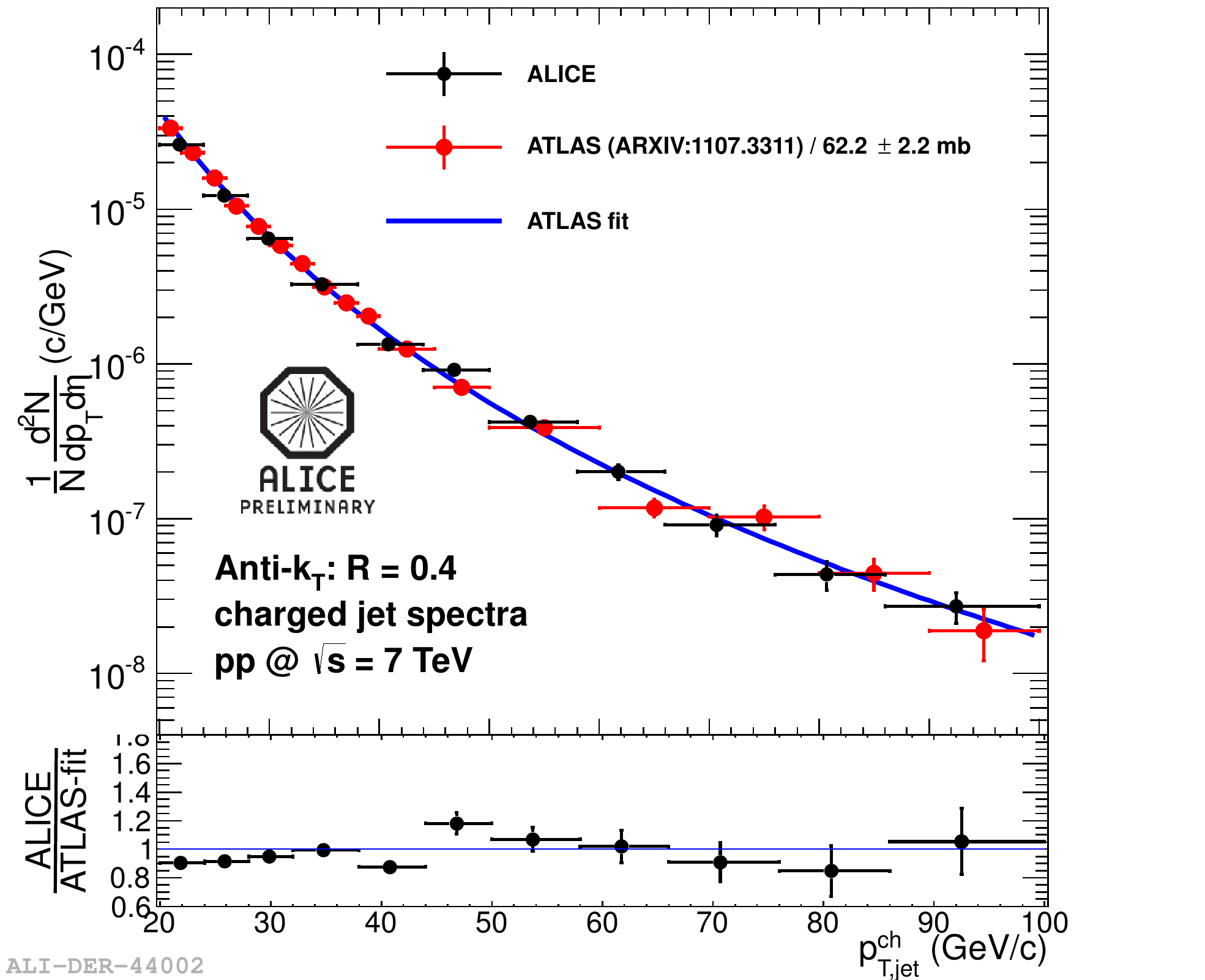} }
        \subfigure[anti-k$_{\rm T}$,~R = 0.6 ] { \label{ATLAS-R06} \includegraphics[width=0.46\textwidth]{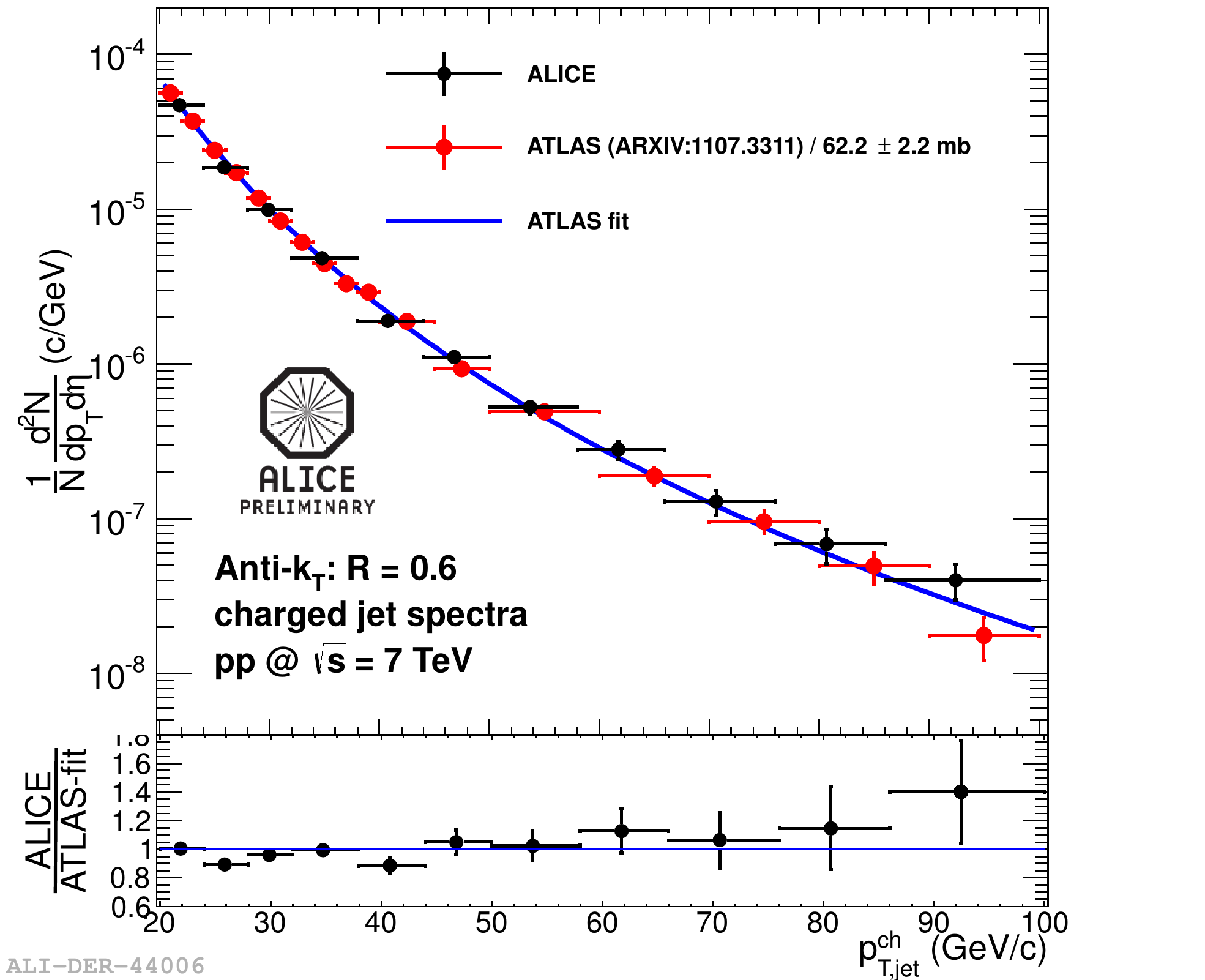} }
        \caption[Figure with comparison of jet spectra to ATLAS result.]{Charged jet spectra measured by the ALICE and ATLAS experiments (black and red points respectively), with minimum transverse momentum of 150~MeV/$c$ and 300~MeV/$c$, respectively,  without underlying event subtraction. The blue line represents Tsallis' fit of ATLAS data. The lower panels show comparisons of the ALICE datapoints to the ATLAS data fit. Fig.~\ref{ATLAS-R04} shows anti-k$_{\rm T}$~jet spectra reconstructed with resolution parameter of 0.4 and in Fig.~\ref{ATLAS-R06} jet spectra with resolution parameter value of 0.6. }
        \label{ATLAS-comp}
      \end{center}
    \end{figure}

      In order to reconstruct jets, we utilize reconstructed tracks as input for FastJet's~\cite{fastjet} anti-k$_{\rm T}$~\cite{antikt} algorithm, sequentially merging particles into proto-jets according to their transverse momenta and distance from each other. The anti-k$_{\rm T}$ algorithm starts merging with particles with the highest transverse momenta. 
  We reconstruct jets with various resolution parameters $R$, a measure of the size of the jet in the space of pseudo-rapidity~($\eta$) and azimuthal angle~($\varphi$)~\footnote{$R$ is measure of distance in space of pseudo-rapidity and azimuthal angle and distance is calculated as $r = \sqrt{\left( \eta_i - \eta_j \right)^2 - \left( \varphi_i - \varphi_j \right) ^2}$ }, ranging from 0.2 to 0.6.  The jet $eta$ acceptance was reduced by the value of this parameter in order to exclude partially reconstructed jets from the analysis.

    \subsection{Corrections}
      Due to the composite nature of colliding protons, more simultaneous (mostly soft) processes may occur in one collision, the so called \emph{underlying event}. After jet reconstruction, we have to subtract effects of these simultaneous processes. This was obtained on event by event basis from tracks in cones perpendicular to the jet axis in azimuth and at the same pseudo-rapidity. Then, the jet spectra are corrected for effects of detector reconstruction, namely track momentum resolution and efficiency of particle reconstruction. 
    This efficiency is the main source of systematic uncertainties of the jet spectra, increasing with the jet transverse momentum. Other sources of uncertainty are the track momentum resolution and the unfolding method, with contribution of few per~cents. Moreover, in case of jet shapes and fragmentation observables, non-negligible uncertainties and corrections are due to the presence and subtraction of underlying event and contamination from secondary decays. 
    For the correction of jet spectra a Bayesian unfolding method~\cite{Bayes,RooUnfold} was used and cross checked with results from Singular Value Decomposition of the response matrix~\cite{unfold-svd}. Jet shapes and fragmentation observables are corrected using a bin-by-bin method, where data are corrected with factors obtained from Monte Carlo simulations by reconstructing the same observable at generator and detector level.

    \begin{figure}[t]
    \begin{center}
      \centering
      \subfigure[ratios of jet cross sections]{\label{r-ratio} \includegraphics[width=0.44\textwidth]{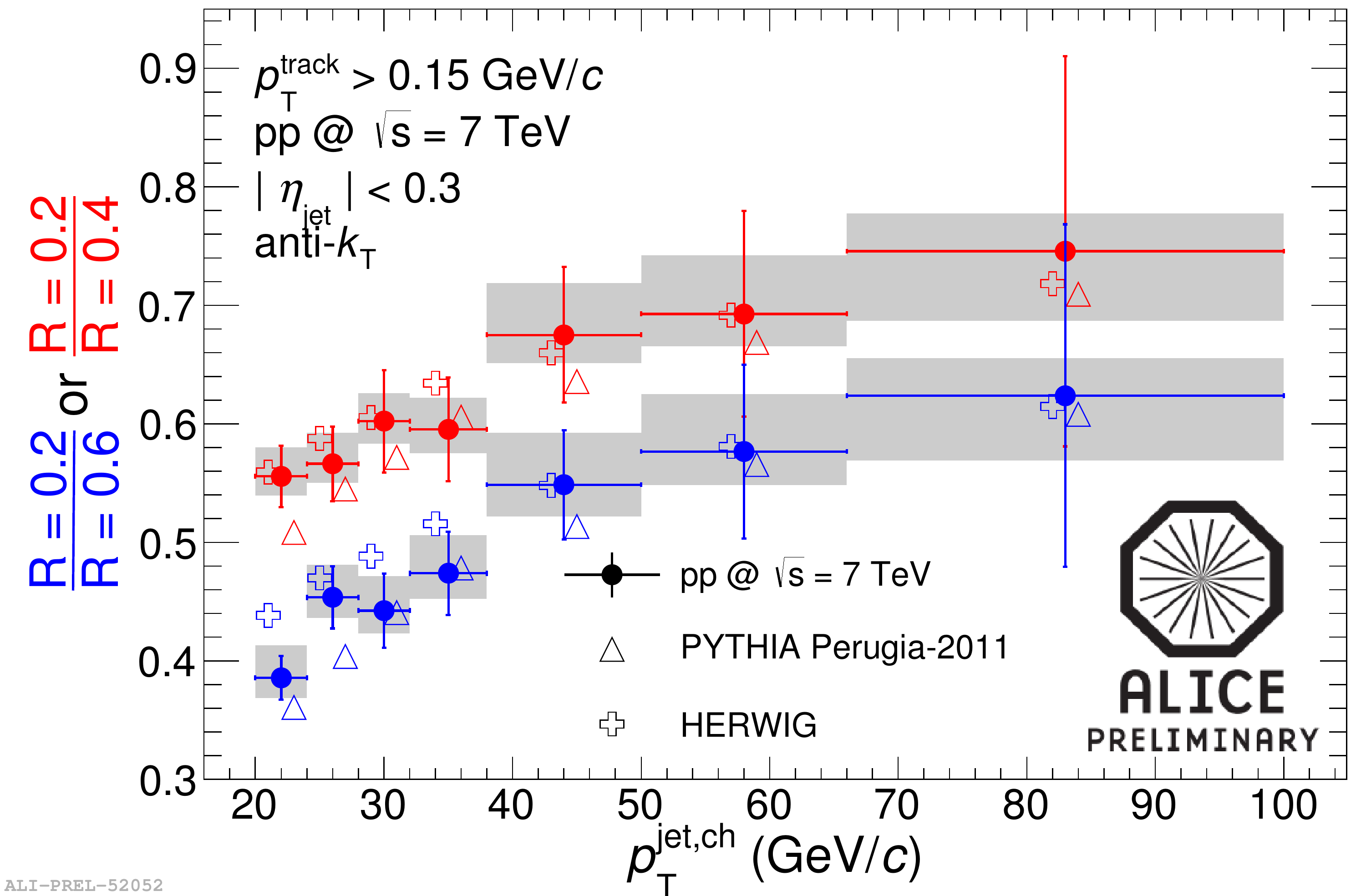} }
      \\
      \subfigure[$p_{\rm T}^{\rm{sum}}$ density for $20 < p^{\rm{ch,jet}}_{\rm T} < 30$ GeV/$c$]{\label{lowptbin-ptdistinjet} \includegraphics[width=0.44\textwidth]{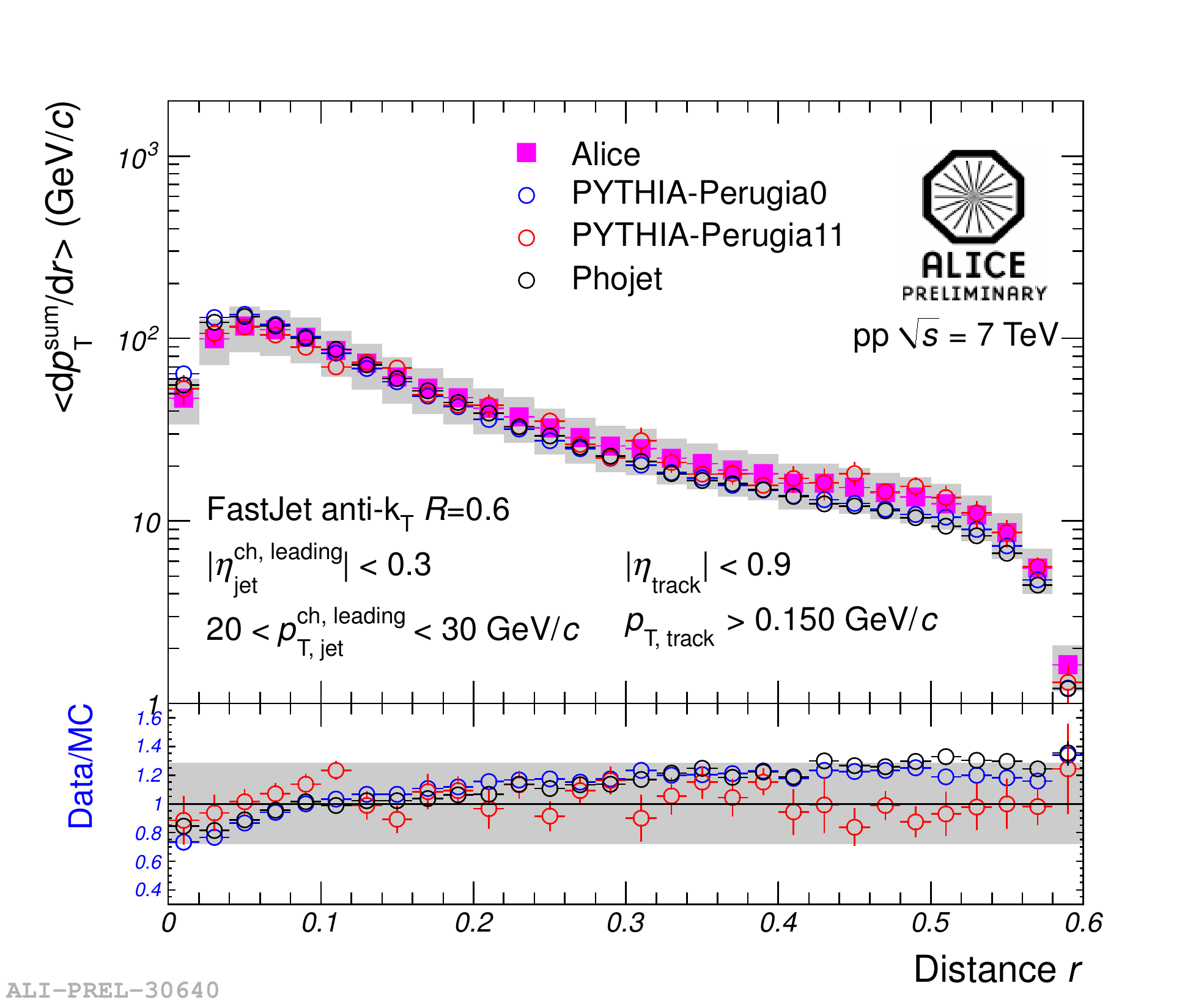} }
      \subfigure[$p_{\rm T}^{\rm{sum}}$ density for $60 < p^{\rm{ch,jet}}_{\rm T} < 80$ GeV/$c$]{\label{highptbin-ptdistinjet} \includegraphics[width=0.44\textwidth]{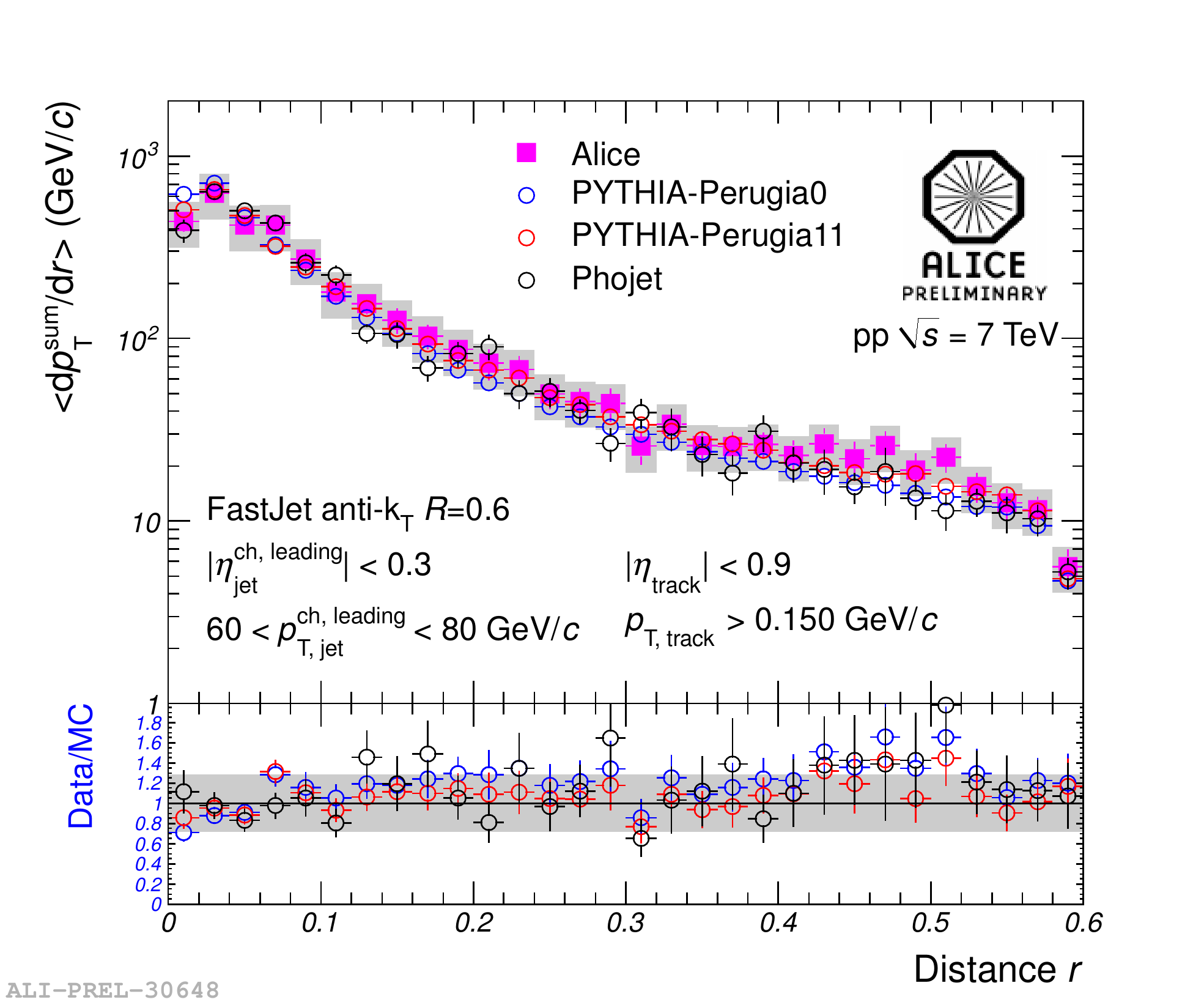} }
      \caption[Figure with jet collimation plots.]{Fig.~\ref{r-ratio} shows ratios of jet spectra reconstructed with resolution parameter $0.2$ and $0.4$ (full red circles), and between spectra with resolution parameters $0.2$ and $0.6$ (full blue circles). Comparison to Monte Carlo simulations is made using HERWIG and PYTHIA Perugia-2011 (open crosses and triangles, respectively). Fig.~\ref{lowptbin-ptdistinjet} and Fig.~\ref{highptbin-ptdistinjet} show summed transverse momentum distribution inside jets as a function of distance $r$ from jet axis, for jets in 20-30~GeV/$c$~and 60-80~GeV/$c$~$p_{\rm T}$~bins, respectively. In lower panel comparison between data (full magenta squares) and predictions from PYTHIA tunes Perugia-0 and Perugia-2011 and PHOJET (open blue, red and black circles respectively) are shown. }
      \label{collimation-plot}
    \end{center}
    \end{figure}

  \section{Results}
    \subsection{Jet spectra}
    Jet spectra were reconstructed from $1.9\times 10^8$ minimum bias collisions with centre of mass energy of 7~TeV and $7 \times 10^7$  minimum bias collisions at 2.76~TeV. These spectra are shown in Fig.~\ref{jet-spectra}, with statistical uncertainties represented with error bars and systematic uncertainties shown as gray boxes. 

    We compare our results with predictions from Monte-Carlo generators, Fig.~\ref{mc-comp}. Considering several different tunes of the PYTHIA~\cite{pythia, PerugiaTunes} generator, Perugia-0, Perugia-2010, Perugia-2011 and AMBT1~\cite{AMBT1}, and the  HERWIG~\cite{herwig, herwig2} generator. None of the generators provides a satisfactory description. However, these improve with increasing transverse momentum. For 7~TeV data, the best description is provided with the HERWIG for low and the PYTHIA Perugia-2011 for high transverse momenta. 

    A comparison of the measurements with results obtained by the ATLAS experiment~\cite{ATLAS-results} is shown in Fig.~\ref{ATLAS-comp}. No underlying event subtraction was done for those results, therefore we produced spectra without subtracted underlying event as well. Moreover, the ATLAS analysis uses tracks with transverse momenta of at least 300~MeV/$c$. Although no correction for this difference was done, we observe a good consistency of the ALICE and ATLAS measurements. This is due to a small influence of different track cuts on final spectra. 

  \begin{figure}[t]
    \begin{center}
      \subfigure[track $p_{\rm T}$~in jets with $ 20 < p_{\rm T} < 30$ GeV/$c$]{\label{trackpt-lowptbin} \includegraphics[width=0.47\textwidth]{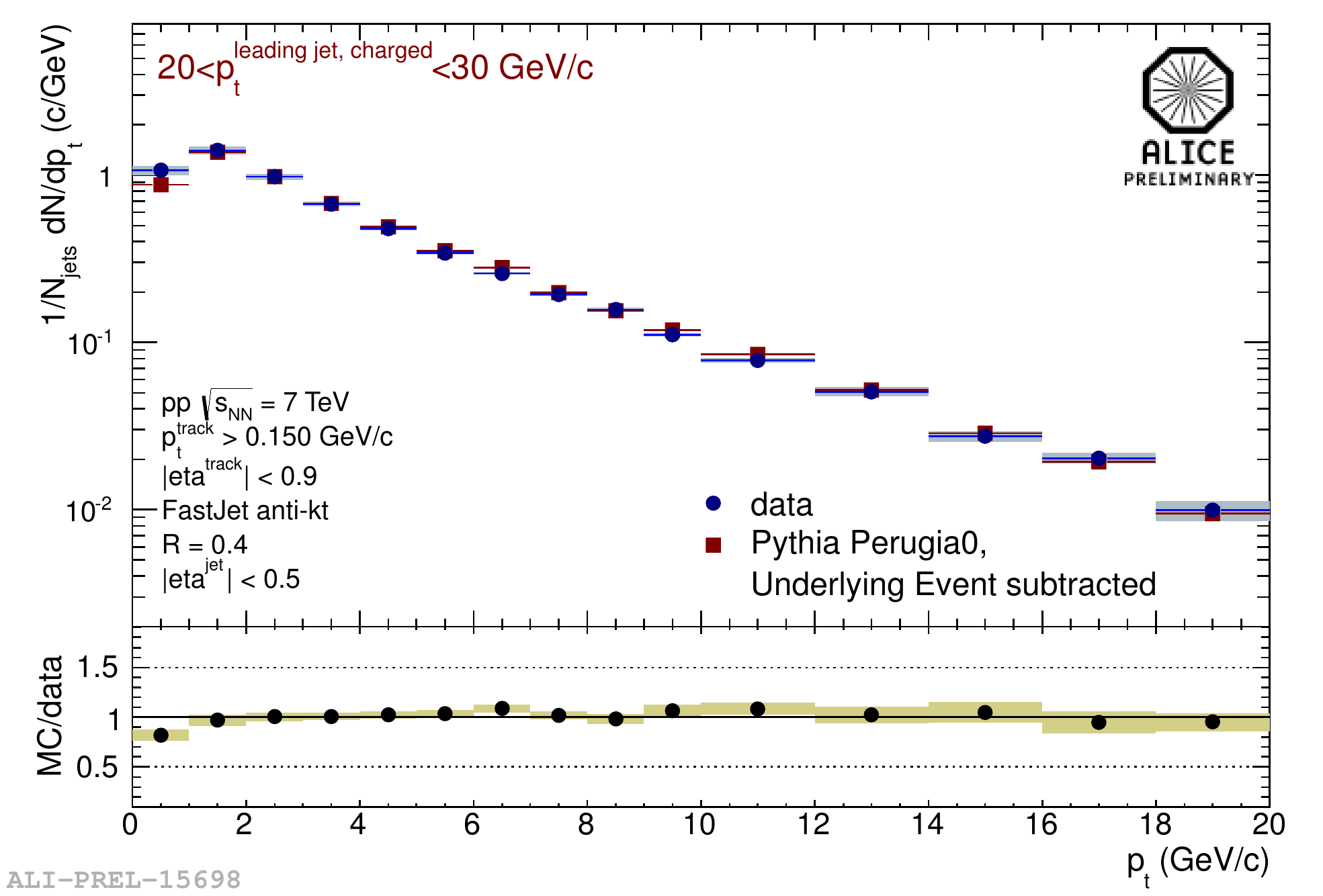} }
      \subfigure[track $p_{\rm T}$~in jets with $ 60 < p_{\rm T} < 80$ GeV/$c$ ]{\label{trackpt-highptbin} \includegraphics[width=0.47\textwidth]{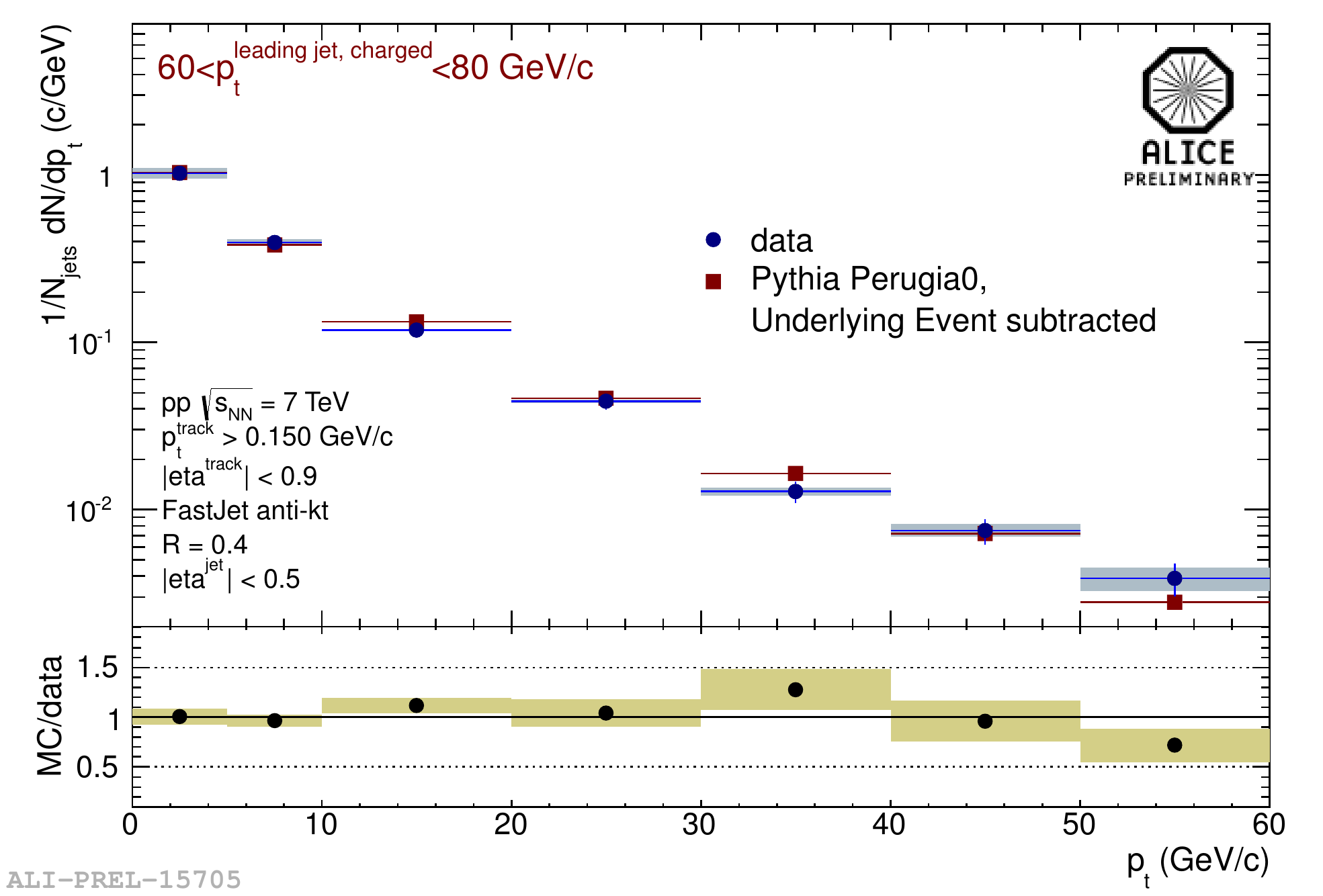} }
      \\
      \subfigure[$\xi$ for jets with $ 20 < p_{\rm T} < 30$ GeV/$c$]{\label{scaled-xi-lowpt} \includegraphics[width=0.47\textwidth]{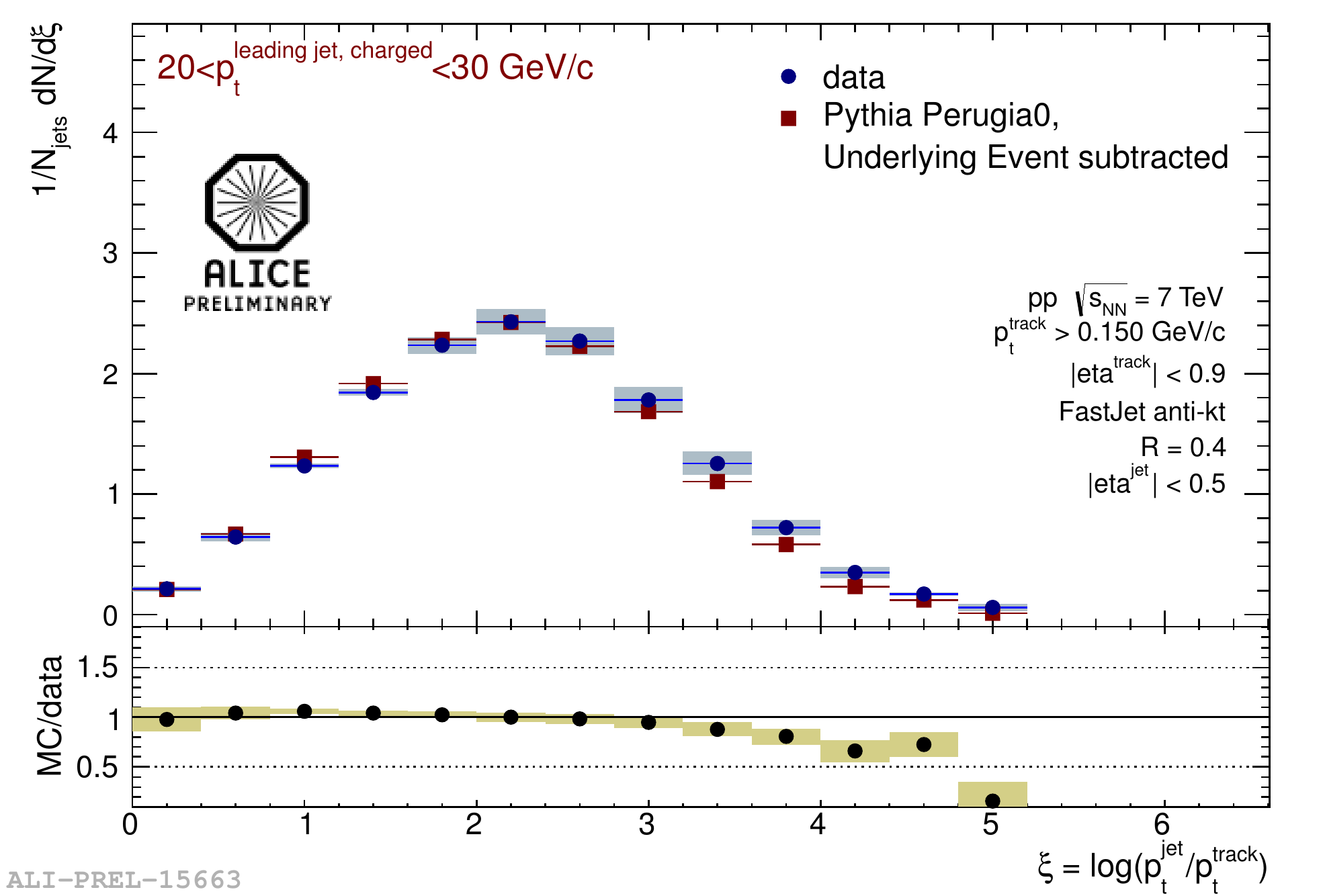} }
      \subfigure[$\xi$ for jets with $ 40 < p_{\rm T} < 60$ GeV/$c$]{\label{scaled-xi-highpt} \includegraphics[width=0.47\textwidth]{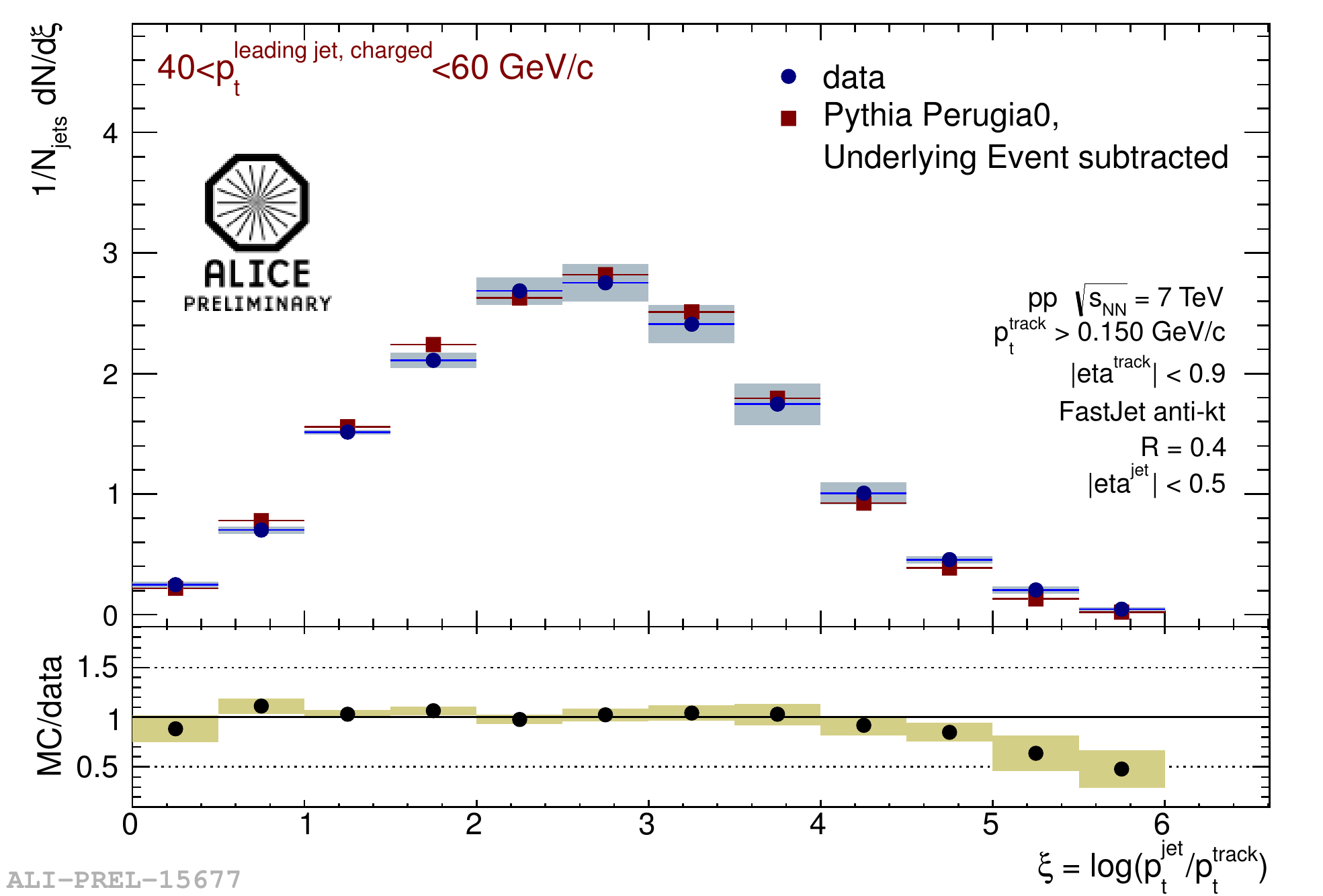} }

      \caption[Figure with jet longitudinal properties.]{Transverse momentum distributions of jet constituents, Fig~\ref{trackpt-lowptbin} and Fig.~\ref{trackpt-highptbin}, and transverse momentum distribution scaled to transverse momentum of jets, Fig.~\ref{scaled-xi-lowpt} and Fig.~\ref{scaled-xi-highpt}, for two jet transverse momentum bins of anti-k$_{\rm T}$~jets with resolution parameter of 0.4.}
      \label{longitudinal-properties-of-jets}
    \end{center}
  \end{figure}

    \subsection{Jet properties}

    Jet fragmentation was studied using various observables that address transverse and longitudinal momentum  distributions with respect to the jet axis. 
    
    To the first category belongs the ratio of jet spectra with different resolution parameters and transverse momentum distribution around the jet axis.  The comparison of jet spectra reconstructed using different resolution parameters shows, Fig.~\ref{r-ratio}, increasing ratio with increasing transverse momentum and a similar behaviour is observed in predictions from Monte Carlo simulations.

    Increasing ratio indicates a stronger collimation of jets as larger fraction of jet momentum is reconstructed by jetfinder with smaller resolution parameter. Ratio of 1 would indicate, that jets reconstructed with different resolution are completely contained within area defined by smaller resolution, and no additional energy is reconstructed with jetfinder with larger resolution.  Although this is an indirect observation of jet collimation, same observable can be obtained from studies of heavy ions collisions. 
    
    The transverse momentum density of charged tracks is other observable quantifying increased collimation of particles within the jet. In Fig.~\ref{lowptbin-ptdistinjet} and Fig.~\ref{highptbin-ptdistinjet} we show a comparison between transverse momentum distributions of jets as a function of distance from jet in two different jet $p_{\rm T}$~bins, 20-30 GeV/$c$~and 60-80 GeV/$c$. From these distributions, increased transverse momentum density of jets at small $r$  for higher jet $p_{\rm T}$~bin is observed. This is  consistent with predictions from Monte Carlo simulations. 

  Concerning the longitudinal properties of jets, we focus on transverse momentum distributions of jet particles in two jet $p_{\rm T}$~bins, Fig.~\ref{trackpt-lowptbin} and Fig.~\ref{trackpt-highptbin}, show a small dependence on jet energy. In Fig.~\ref{scaled-xi-lowpt} and Fig.~\ref{scaled-xi-highpt} scaled transverse momenta distributions of these tracks with respect to the transverse momentum of the jet they belong to, $\xi = \log \left( p_{\rm T}^{\rm{ch, jet}} / p_{\rm T}^{\rm{ch, track}}\right)$. Here, the distribution of $\xi$ is shown for two different jet $p_{\rm T}$~bins, 20-30~GeV/$c$~and 40-60~GeV/$c$. The characteristic shape of distribution, ``hump-backed'' plateau, is observed, which is a result of QCD coherence~\cite{QCD coherence}. 
  The position of this plateau moves to higher values of $\xi$ and the area under the distribution increases, with increasing transverse momentum of jets, as expected from QCD calculations based on Modified Leading Logarithmic Approximation~\cite{MLLA}. Our observations are quantitatively consistent with predictions from Monte Carlo simulations obtained with the PYTHIA.

  \section{Summary and Conclusions}
  We have analysed jet spectra from proton--proton collisions at centre-of-mass energies of 2.76~TeV and 7~TeV with the anti-k$_{\rm T}$~algorithm using several different jet resolution parameters. Large variations in descriptions of cross sections provided by various tunes of Monte Carlo generators are observed. On the other hand, Monte Carlo predictions are consistent with our measurement of jet shape observables and fragmentation distributions from proton--proton collisions at centre of mass energy of 7~TeV. The \emph{hump-backed plateau} is observed in the $\xi$ distribution as predicted by QCD coherence and its behaviour is in a qualitative agreement with theoretical calculations. Our observation of momentum dependence of jet collimation is consistent with Monte Carlo predictions within uncertainties for both direct and indirect measurements. 

  \section*{Acknowledgement}
  This work was supported by Grant Agency of the Czech Technical University in Prague, grant SGS13/215/OHK4/3T/14, and by Ministry of Education of the Czech Republic, grant LG 13031.

\end{document}